\pdfoutput=1

\documentclass[iop,apj]{emulateapj}

\usepackage{natbib}
\bibpunct{(}{)}{;}{a}{}{,}
\usepackage{amsmath}
\usepackage{amsfonts}
\usepackage{color}
\usepackage{graphicx}

\shorttitle{Cosmic Fullerenes from Arophatic Clusters}
\shortauthors{Micelotta et al.}

\begin{document}


\title{The Formation of Cosmic Fullerenes from Arophatic Clusters}


\author{Elisabetta R. Micelotta\altaffilmark{1}}
\affil{Department of Physics and Astronomy, University of Western Ontario, \\
  London, Ontario N6A 3K7, Canada}
\email{email: emicelot@uwo.ca}
\author{Anthony P. Jones\altaffilmark{2}}
\affil{Institut d'Astrophysique Spatiale,  CNRS/Universit{\'e} Paris Sud, 91405 Orsay, France}
\author {Jan Cami\altaffilmark{1, 3} and Els Peeters\altaffilmark{1, 3}}
\affil{Department of Physics and Astronomy, University of Western Ontario, \\
  London, Ontario N6A 3K7, Canada}
\altaffiltext{3}{SETI Institute,189 Bernardo Avenue, Suite 100, Mountain View, CA 94043, USA}
\author{Jeronimo Bernard-Salas\altaffilmark{2}}
\affil{Institut d'Astrophysique Spatiale,  CNRS/Universit{\'e} Paris
  Sud, 91405 Orsay, France}
\and
\author{Giovanni Fanchini\altaffilmark{1}}
\affil{Department of Physics and Astronomy, University of Western Ontario, \\
  London, Ontario N6A 3K7, Canada}


\begin{abstract}
Fullerenes have recently been identified in space and they may play
a significant role in the gas and dust budget of various astrophysical
objects including planetary nebulae (PNe), reflection nebulae (RNe)
and H{\sc ii} regions.  The tenuous nature of the gas in these
environments precludes the formation of fullerene materials following
known vaporization or combustion synthesis routes even on astronomical
timescales.  We have studied the processing of hydrogenated amorphous
carbon (a-C:H or HAC) nano-particles and their specific derivative
structures, which we name ``arophatics'', in the circumstellar
environments of young, carbon-rich PNe.  We find that UV-irradiation
of such particles can result in the formation of fullerenes,
consistent with the known physical conditions in PNe and with
available timescales.
\end{abstract}

\keywords{circumstellar matter -- infrared: general -- ISM: molecules 
-- planetary nebulae: general -- stars: AGB and post-AGB}



\section{Introduction}\label{intro}

The gas and dust that is ejected by dying stars and from which new
stars will form, is constantly being processed chemically as well as
physically \citep[{\it e.g.},][]{tielens05}. A particularly rich and
diverse route is followed by carbon. In the cool surroundings of
carbon stars -- evolved stars that have dredged up large amounts of
carbon to the stellar surface -- more than 60 individual molecular
species have been identified, including benzene, polyynes and
cyanopolyynes \citep{kwok09,cernicharo01, pardo07}. In addition, the
spectra reveal diverse dusty materials such as hydrogenated amorphous
carbon (HAC or a-C:H) and silicon carbide (SiC). As the mass-loss
process strips the outer layers of the star, a hot ($T \ge 30,000$ K)
central white dwarf becomes exposed, whose strong UV irradiation
further processes the ejecta and makes them visible as a planetary
nebula (PN). A key component of the carbon dust inventory in PNe are
polycyclic aromatic hydrocarbons (PAHs) and PAH-like species, a class
of large and hardy carbonaceous species whose formation mechanisms are
unclear \citep{cherchneff00}. Although not a single PAH member has
been identified in space, their characteristic spectral features at
infrared wavelengths are observed throughout the Universe, from which
it is inferred that they reside ubiquitously in space and could lock
up as much as 15\% of the cosmic carbon \citep[{\it
e.g.},][]{tielens05}.

Two other large aromatic species have been recently identified in
space: C$_{60}$ and C$_{70}$. These are the best-known members of the
family of fullerenes, a class of molecules made of hexagonal and
pentagonal aromatic carbon rings, fused in the shape of a hollow
sphere or ellipsoid. The most abundant of these molecules, the
Buckminsterfullerene C$_{\rm 60}$, has the structure of an old-fashioned
black and white soccer ball. The near-spherical carbon configuration
of these two molecules, which have a closed surface, a closed-shell
electron distribution and an almost unstrained network, results in a
very high stability against dissociation and prolonged exposure to
high temperature.  Fullerenes, in particular C$_{\rm 60}$, have
peculiar and appealing photochemical, electrochemical and physical
properties, which can be exploited in various fields, from
nanotechnology to medicine, to space science.

C$_{\rm 60}$ and C$_{\rm 70}$ were discovered during laser ablation
experiments on graphite targets, aiming to study long carbon chains in
interstellar clouds \citep{kroto85}. Because of their remarkable
stability, fullerenes appeared particularly suited to survive the
harsh conditions of the interstellar medium. Their unique properties
indicate that they may play an important role in organic
astrochemistry and astrobiology.  Cosmic fullerenes remained elusive
until the recent discovery of C$_{\rm 60}$ and C$_{\rm 70}$ in the
planetary nebula Tc 1 by \citet{cami10}.  Fullerenes have since been
confirmed in many more evolved star environments
\citep{garcia10,garcia11b,zhangkwok11,clayton11,gielen11,roberts12,evans12},
as well as in young stellar objects \citep{roberts12}, reflection
nebulae (RNe) \citep{sellgren10, peeters12} and photodissociation regions
\citep{rubin11}.

Fullerenes can be efficiently synthesized in the laboratory by the
vaporization of carbon rods in an electric arc
\citep{kratschmer90a,kratschmer90b} and from hydrocarbon combustion
under optimised conditions \citep{howard91,nanoC04}. However, the
formation routes of fullerenes in space are still unknown.  We review
the currently known formation mechanisms of fullerenes in the
laboratory (\S~\ref{earth_sec}), showing why these methods would not
work in space (\S~\ref{not_working_sec}).
We propose an alternative pathway, consistent with astrophysical
conditions, based on the photo-processing of a family of carbonaceous
species which we name ``arophatics'' (\S~\ref{our_mechanism_sec}). We
discuss the observations of fullerenes in PNe (\S~\ref{observations_sec}) and 
finally we summarize our conclusions (\S~\ref{conclusions_sec}).

\section{Fullerene formation on Earth}\label{earth_sec}

\subsection{The discovery of fullerene}\label{kroto_sec}

The original discovery of C$_{\rm 60}$ and C$_{\rm 70}$ dates back to
1985 with the experiment of \citet{kroto85}. In this experiment,
carbon species were vaporized from a graphite target into a He flow
with tunable pressure/density. The vaporization was done by a pulsed
laser, the resulting carbon clusters were expanded in a supersonic
molecular beam, photoionized and their masses measured by a
time-of-flight mass spectrometer.  Vaporization at low He pressure
(less than 10 torr) leads to a broad distribution of clusters, with 38
to 120 carbon atoms (always even numbers), with C$_{\rm 60}$ and
C$_{\rm 70}$ present but not the dominant species.  At a pressure of
760 torr the C$_{\rm 60}$ and C$_{\rm 70}$ peaks clearly
dominated. When the expansion of the He+carbon cluster mixture was
delayed, the resulting mass distribution was completely dominated by
C$_{\rm 60}$ - 50\% of total large cluster abundance, and C$_{\rm
70}$, with 5 \% of total large cluster abundance. These distributions
have been interpreted as due to the increasing number of He-cluster
and cluster-cluster collisions, resulting in the ``survival of the
fittest'', i.e. the more stable clusters. The famous icosahedral
soccer ball structure was proposed for the first time for C$_{\rm
60}$, and later confirmed by spectroscopic studies
\citep{kratschmer90a,kratschmer90b,taylor90}. However, the elementary
reaction mechanisms occurring during the ``thermalization'' process
remained unknown.

\subsection{The role of temperature: mutually exclusive formation
of PAHs and fullerenes}\label{temperature_sec}

\citet{jager09} experimentally studied the gas-phase formation of
carbonaceous compounds. Their first experiment was essentially a
replication of the Kroto experiment, i.e.  laser vaporization of a
graphite target in a quenching atmosphere of He or He/H$_2$ with
pressures between 3.3 and 26.7 mbar. The vibrational temperature of
the laser-induced plasma was estimated to be between 4000 and 6000 K
for the laser power densities used.  In a second set of experiments,
the laser-induced pyrolysis of C$_2$H$_2$ and C$_6$H$_6$ using a
pulsed laser with high power densities was studied. The resulting
condensation temperatures were above 3500 K, comparable with the
vaporization experiment.  In a third set of experiments, laser-induced
pyrolysis of the same hydrocarbon precursors was carried out, but this
time at a much lower temperature (max 1700 K).

The analysis of the condensation products showed a striking effect of
the temperature. During the high temperature experiments, only
fullerene-like soot and fullerenes were produced. During the low
temperature experiments, only soot and PAHs were formed (100 \% PAHs
at $\sim$ 1000 K). 
The results clearly tell that the temperature determines which kind of
condensates will be formed. Moreover, the two pathways seem to be
mutually exclusive, at least under these experimental conditions:
fullerenes and PAHs cannot be formed together. It is important to note
that high pressure has been used during the experiments to concentrate
the condensation within a small volume, i.e. to mantain a high density
of the condensing species.

\subsection{Molecular dynamics simulations of fullerene formation}\label{simulation_sec}

The laboratory experiments described in the previous sections do not
tell us how fullerene is formed from graphite and hydrocarbons, but
only under which conditions this happens. The elementary reaction
mechanisms involved in the formation of fullerenes can be investigated
performing quantum chemical molecular dynamics simulations
\citep{zheng05, irle06, zheng07}. In one set of simulations a hot
carbon vapor was reproduced by putting 40 randomly oriented C$_2$
molecules in a tiny 20\,\AA\ cubic box \citep{zheng05,
zheng07}. The system was let to evolve, and extra C$_2$ molecules
were added at fixed times in order to simulate an open
environment. During the first stage, the system was kept at a constant
temperature of 2000 K and giant carbon cages self-assembled. During
the second stage of the simulation the temperature was raised to 3000
K, and the shrinking of these hot giant fullerenes down to C$_{\rm
65}$ was observed  \citep{zheng07}.

The formation of the giant cages starts with nucleation of polycyclic
structures (hexagons and pentagons) from entangled polyyne chains.
This is followed by growth of the structure and finally cage closure
similar to the self-capping of open-ended single-walled
nanotubes. Because of the rapid gain of energy due to cage closure,
the giant fullerenic cages are produced in a vibrationally highly
excited state. The excess energy has to be dissipated, either by
unimolecular dissociation or collision with other carbon clusters or
carrier gas atoms. The simulations show that the newly produced giant
fullerenes inevitably shrink to smaller sizes.  All the road maps to
fullerene formation proposed by previous models were associated with
intermediate structures that are in thermodynamic equilibrium, while a
hot carbon vapor is indeed a system far from thermodynamic
equilibrium.  In the model developed by Irle, Morokuma and
collaborators, for the first time the dynamic self-assembly of
fullerene molecules occurs as an irreversible process emerging
naturally under the nonequilibrium conditions typical of a hot carbon
vapor.

\subsection{Direct formation of fullerenes from graphene}\label{chuvilin_sec}

The experimental results of \citet{ugarte92} have presented
evidence of the spontaneous tendency of electron-irradiated graphite
to include pentagons in its hexagonal network, hence form curved
structures. This has been further confirmed by \citet{chuvilin10}, who
have shown that fullerenes can form directly from graphene in a
similar fashion.  In the experiment of Chuvilin and co-workers, a
sheet of graphene was exposed to a beam of energetic 80 keV
electrons. The energy transferred to the graphene results in the loss
of carbon atoms from the edges of the sheet. If the size of the
graphene flake is not too big (less than a few hundreds of carbon
atoms), the loss of carbon atoms at the edge will lead to the
formation of pentagons, which triggers the curving of graphene into a
bowl-shaped structure. Further carbon removal from the edges using the
electron beam will reduce the size of the curved structure until it is
sufficiently small to zip up its open edges and isomerize into a
closed fullerene structure.

\subsection{Formation of fullerenes via closed network growth}\label{CNG_sec}

\citet{dunk12} claim having identified  the fundamental processes
leading to the formation of fullerenes in a recent experimental
work, based on laser vaporization of pure C$_{\rm 60}$ in the presence
of carbon vapor. According to this study, fullerenes would
self-assemble bottom-up through a closed network growth (CNG)
mechanism based on the ingestion of atomic carbon and C$_2$
clusters. It should be noted that the work from \citet{dunk12}
provides experimental evidence of the growth of larger fullerenes from
pre-existing C$_{\rm 60}$ only.  Because the experiments show that the
CNG of larger fullerenes does not result in the production of C$_{\rm
60}$, it is deduced that C$_{\rm 60}$ formation must be a result of
CNG from smaller fullerenes. However, the initial formation mechanism
of such smaller fullerenes is still under debate.

\section{Fullerene formation in space: why terrestrial 
methods do not work}\label{not_working_sec}

Although all the chemical ingredients and the required temperatures
for graphite vaporization and hydrocarbon combustion can be found in
astrophysical environments, the densities are far too low to proceed
to fullerene formation on reasonable timescales.  In space, the most
favourable conditions for fullerene formation via
vaporization/combustion are found in a post-shock gas.  In such
environments the fullerene building blocks, i.e.  C$_{2}$ groups and
polycyclic species coming from the fragmentation of PAHs
\citep{micelotta10b, micelotta10a}, ought to exist in the required
vibrationally excited states.  However, the post-shock carbon
densities are low and represent a serious obstacle to fullerene
formation. We have derived the scaling rule
(Eq.~\ref{scaling_eq}) relating the time $\tilde{\tau}$, required for
fullerene condensation, with the initial density $n$ of the carbon
gas. The details of the calculation are reported in the Appendix. This
scaling rule has then been used to estimate the time necessary for the
coagulation of fullerenes from a carbon gas with interstellar
densities ($n \sim$ 10$^{-4}$~carbon~cm$^{-3}$). The scaling rule is:
\begin{equation}
\label{scaling_eq}
  \frac{\tilde{\tau_2}}{\tilde{\tau_1}} = \frac{n_1^2}{n_2^2}\,\,
  \Rightarrow \,\, \tilde{\tau_2} = \frac{\tilde{\tau_1} n_1^2}{n_2^2},
\end{equation}
where $\tilde{\tau_1}$ and $\tilde{\tau_2}$ are the condensation
timescales corresponding to densities $n_1$ and $n_2$ respectively.
Knowing the density and timescale in simulations, $n_1$ and
$\tilde{\tau_1}$, and the density of the interstellar gas, $n_2$, 
we can then calculate the fullerene condensation time in space,
$\tilde{\tau_2}$. We adopt $n_1$~=~10$^{22}$~carbon~cm$^{-3}$ and
$\tilde{\tau_1}$~=~50 ps \citep{zheng05}. The
hydrogen density of the post-shock gas is $\sim$~1~cm$^{-3}$, but
carbon is 10$^{-4}$ less abundant, so
$n_2$~=~10$^{-4}$~carbon~cm$^{-3}$. From Eq.~\ref{scaling_eq} we obtain
$\tilde{\tau_2}$~=~1.6~$\times$~10$^{34}$ yr, which is longer than the age of
the Universe. We also estimate the carbon density $n_2$ required to
form fullerenes in a shocked gas before the temperature drops below
1000 K. At this temperature the condensation of fullerenes cannot
occur. For a typical interstellar shock with a velocity of
125~km~s$^{-1}$ the cooling time (after which the temperature of the
post-shock gas drops below 1000 K) is about 2$\times$10$^5$ yr. Under
these conditions, the required carbon atom density for the
condensation of fullerenes is
$n_2$~=~3$\times$10$^{10}$~carbon~cm$^{-3}$, and in space such
densities do not exist outside of stellar atmospheres.

\citet{berne12} propose a scheme for the direct formation of C$_{\rm
60}$ from PAHs in the reflection nebula NGC 7023 based on graphene
dissociation experiments from \citet{chuvilin10}. In NGC 7023
fullerenes and PAHs appear to coexist in the same location close to
the central star, while only PAHs are detected further away from the
star. According to \citet{berne12}, a similar phenomenon to the one
described by \citet{chuvilin10} occurs in NGC 7023, but induced by
photo processing instead of electron bombardment.
The absorption of UV photons from the central star first
completely dehydrogenates the PAHs \citep{tielens05}, producing
graphene flakes.  Further photon absorption induces the loss of
carbons from the edges of the graphene sheet. This, according to
\citet{chuvilin10}, gives rise to pentagonal defects which will induce
the curvature of the sheet and the subsequent closure of the cage to
form a C$_{\rm 60}$ molecule.

There are some issues related to the applicability of the results from
such experiments to interstellar conditions. First of all, it is not
possible in space to tune the number of carbons removed from the
graphene flake in order to get the right size required for cage
closure. The number of ejected carbons depends on the rather slow
photodissociation process, which is governed by the dissociation
parameter $E_0$. This parameter describes the dissociation rate of a
highly excited PAH molecule using an Arrhenius law \citep{tielens05},
and to a first approximation can be seen as the energy that needs to be
concentrated over a bond in order to break it and release a
fragment. The dissociation probability (and hence the dissociation
rate) decreases for increasing $E_0$ because more energy is required
in the bond that has to be broken. The value of $E_0$ has been derived
from experimental data only for very small PAHs (up to 24 C-atoms)
with a very open carbon skeleton. The extrapolation of such results to
larger ($\sim$ 50 C-atoms) and compact astrophysically relevant PAHs
dissociating under interstellar conditions is very uncertain, leading
to values of $E_0$ ranging from 3.65 to 5.6 eV
\citep{micelotta10a}. This corresponds to values for the dissociation
probability going from 0.5 to 3$\times$10$^{-12}$ respectively. To
form fullerenes during the lifetime of NGC 7023, the mechanism
proposed by \citet{berne12} needs a specific tuning of the parameters
involved. Bacause the size of PAHs cannot be controlled as during
experiments, the number of carbons in the precursor molecule is
severely constrained to about 70 atoms, very close to the 60 carbons required to form
C$_{\rm 60}$. More importantly, the most favourable conditions for 
dissociation are chosen, i.e., the lowest value for the dissociation
parameter, $E_0$ = 3.65 eV. However, this choice does not appear physically
motivated. Indeed, if another legitimate value is adopted for the
dissociation parameter, e.g., $E_0$ = 4.6 eV \citep[the value falling in the
middle of the range determined for $E_0$ -- see ][]{micelotta10a}, the
decrease of the dissociation rate is such that the fragmentation
process is far too slow and the conclusion suggested in
\citet{berne12} no longer holds: PAHs with 70 carbon atoms require
longer than the lifetime of NGC 7023 to fragment, therefore they
cannot be the precursors of the C$_{\rm 60}$ observed at the present
epoch.

Thus, none of these mechanisms are feasible in the astrophysical
settings where fullerenes have been detected, and hence, alternative
routes are needed.

\section{Our proposed formation mechanism: from a-C:H to fullerene
through arophatic clusters}\label{our_mechanism_sec}

Trace amounts of fullerenes have been identified
among the dissociation products of UV-irradiated HACs
\citep{1997ApJ...489L.193S,gadallah11}. Together with fullerene
emission, the three PNe Tc 1, SMP LMC 56 and SMP SMC 16 show broad
emission plateaus between 6 -- 9 $\mu$m and between 10 -- 13 $\mu$m,
compatible with emission from H-rich amorphous carbon nanoparticles 
\citep{bsalas12}. This suggests that a viable formation route
in space could exist that starts from HAC processing \citep[{\it
  e.g.},][]{garcia10, garcia11a}.

In this study we focus on the formation of fullerenes in the
circumstellar environments of proto-planetary nebulae (PPNe) and PNe.
HACs are an abundant carbonaceous dust component in both these types
of objects \citep{grishko01,goto03}. In addition, one of the main
characteristics of PNe is their strong radiation field, which will
process the HACs.

\subsection{Arophatic clusters}

In their experiments, \citet{1984JAP....55..764S} and
\citet{1996ApJ...472L.123S} showed that HAC/a-C:H decomposition leads
to a low density material ($\rho \lesssim 1.5$\,g~cm$^{-3}$). In
particular, \citet{1996ApJ...472L.123S} and \citet{1997ApJ...489L.193S}
showed that the decomposition end-product is an aerogel-like material
consisting of weakly-connected aromatic ``proto-graphitic'' or
PAH-like clusters in a ``friable network''.
The dehydrogenation of small carbonaceous particles, containing less
than a few hundred C atoms, will likely lead to structures that
resemble the particles seen by \citet{1997ApJ...489L.193S} in their
experiments and/or the locally aromatic polycyclic hydrocarbons
(LAPHs) proposed by \citet{2003ApJ...594..869P}. 
The particles in the \citet{1997ApJ...489L.193S} experiments are found
to be small alkanes, unsaturated carbon-chain radicals and small
dehydrogenated PAH-like species. After the release of these lighter
molecules \citet{1997ApJ...489L.193S} observed a strong mass peak near
500\,amu ($\simeq 40$ C atom, proto-graphitic, aromatic clusters)
having IR spectra similar to that of HAC/a-C(:H) with absorption and
emission features at 3.3, 3.4 and $6.2\,\mu$m.  The results of these
experiments and recent modelling \citep{jones12b} therefore imply an
intimate mix of aromatic clusters and aliphatic groups in small
carbonaceous species. We note that the LAPH structures studied by
\cite{2003ApJ...594..869P} have compositions in the range
C$_{19}$H$_{22}$ to C$_{36}$H$_{32}$ and contain aromatic domains with
only a few rings per domain, which are linked by bridging aliphatic
ring systems, and should have IR spectral properties similar to
HAC/a-C(:H) materials \citep[{\it
e.g.},][]{jones12c,jones12a,jones12b}.

The conditions required for HAC dehydrogenation are common in space,
we can thus expect the formation of LAPH-like species where the
aromatic components are bridged by olefinic carbon-containing rings
and chains.  The key component in these structures is the presence of
the aliphatic bridging groups linking the aromatic clusters, in what
we here call ``arophatic'' structures.  Such structures, analogous to
an aliphatic periphery on PAH molecules \citep[{\it
e.g.},][]{2008ApJ...679..531R}, could provide sites for molecular
hydrogen formation in PDRs \citep{jones12a}. Size-dependent
$\pi-\pi^\ast$ band considerations indicate that the intrinsic
aromatic clusters are indeed most likely to consist of `isolated' two-
and three-ring systems in such small hydrocarbon particles
\citep{jones12b}.  The non-aromatic carbon atoms are in the form of
``more flexible'' short olefinic and/or aliphatic bridging structures
such as --CH=CH-- and --CH$_2$--CH$_2$--. This generic type of
structure was also proposed by \cite{2000ApJ...543L..85G}, who
suggested that aliphatic cyclo-hydrocarbons are the precursors to
aromatic cluster formation in HAC materials. This kind of
configuration, with aliphatic chains attached to aromatic sites, is
also found in Aromatic-Aliphatic Linked Hydrocarbon (AALH) structures,
which appear to have an important role in the nucleation of
nanoparticles in soot formation in flames \citep{chung11}. It should
be noted that AALHs are small and relatively simple molecules, such as
1,2-di(naphthalen- 2-yl)ethane (DNE), while our proposed ``arophatic''
clusters are bigger and more complex structures.

The aliphatic bridging structures must be quasi-orthogonal to the
aromatic species that they bridge and will be unable to transform into
aromatic structures without compromising ({\it i.e.}, destroying) the
entire structure. These active sites could, likely, interchange
between $sp^2$ olefinic and radical $sp^3$ states with dangling bonds,
in a kind of isomerisation ``flipping'' process, which could play a
key role in cage closure and fullerene formation in the solid state.

The work of \citet{2004PhilTransRSocLondA..362.2477F} indicates that
small clusters within a-C(:H) materials are chain-like, while the
larger clusters are three-dimensional, cage-like, $sp^2$ structures
and as \citet{chuvilin10} show these cage-like structures have a
natural tendency to close-up and form fullerene-like and fullerene
structures.

\begin{figure} [b]
\resizebox{\hsize}{!}{\includegraphics[angle=90]{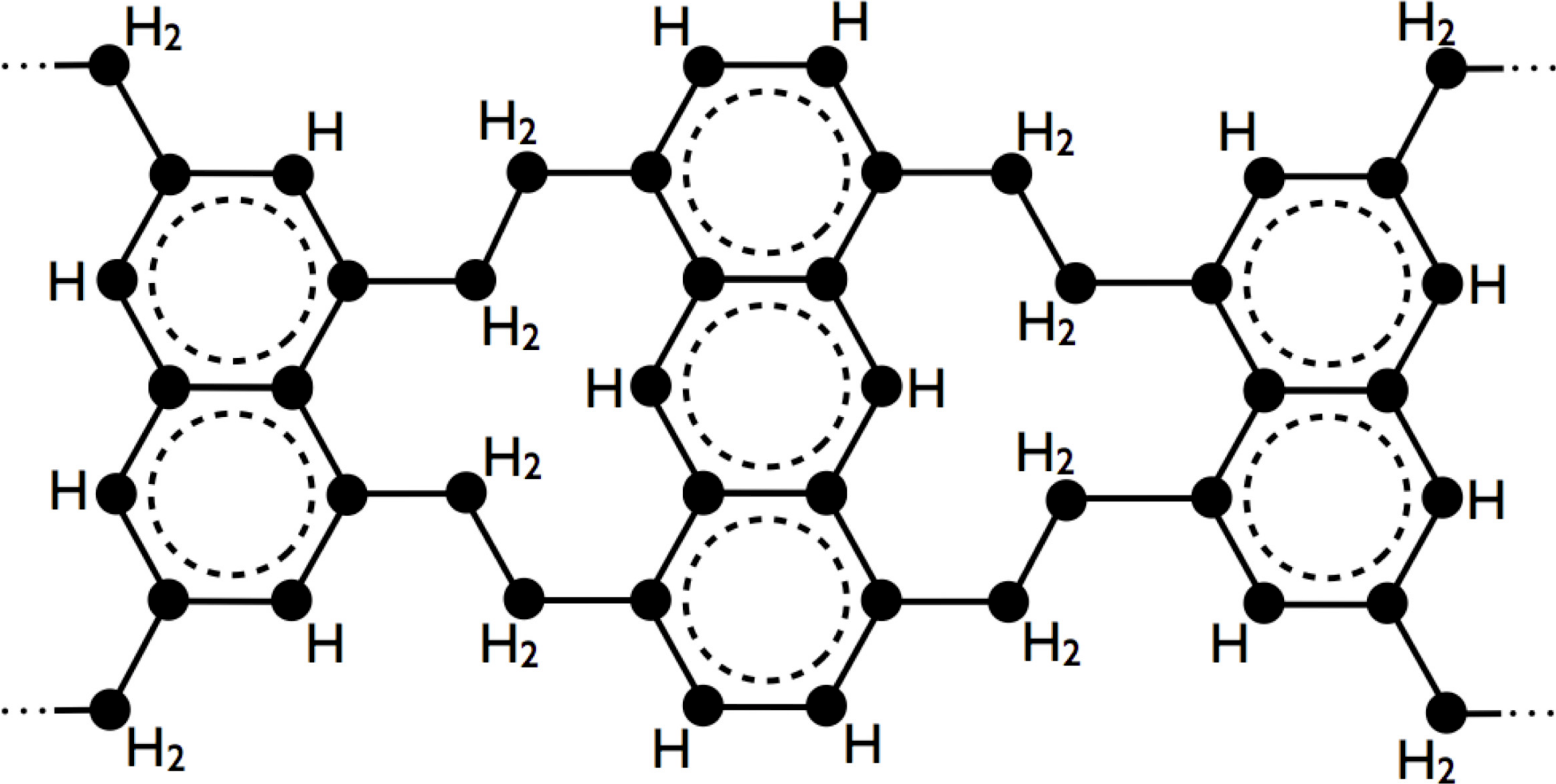}}
 \caption{An example of an un-wrapped small ``arophatic'' cluster
   (C$_{46}$H$_{38}$). The filled circles represent carbon and the dotted 
   bonds indicate the structure-wrapping connections.}
 \label{fig_arophatics_0}
\end{figure}

The  ``arophatic'' clusters are structures evolving from
larger, amorphous hydrocarbon particles and, from a topological point
of view, they will inevitably end up with a tube-like, cup-like or
cage-like appearance \citep[{\it e.g.}, see the two right hand species
in Fig.~1 of][]{2003ApJ...594..869P}.  Such structures are `curled-up'
or `folded-over', defected graphite (DG) or graphene sheets
\citep{jones12c}, and we can therefore represent them by their
`flattened-out' equivalents as per Fig.~12 in \citet{jones12c} but
with aromatic domains reduced in size to only a few rings per aromatic
cluster (Fig.~\ref{fig_arophatics_0}).
%
\begin{figure}
\centering
 \includegraphics[width=0.9\hsize]{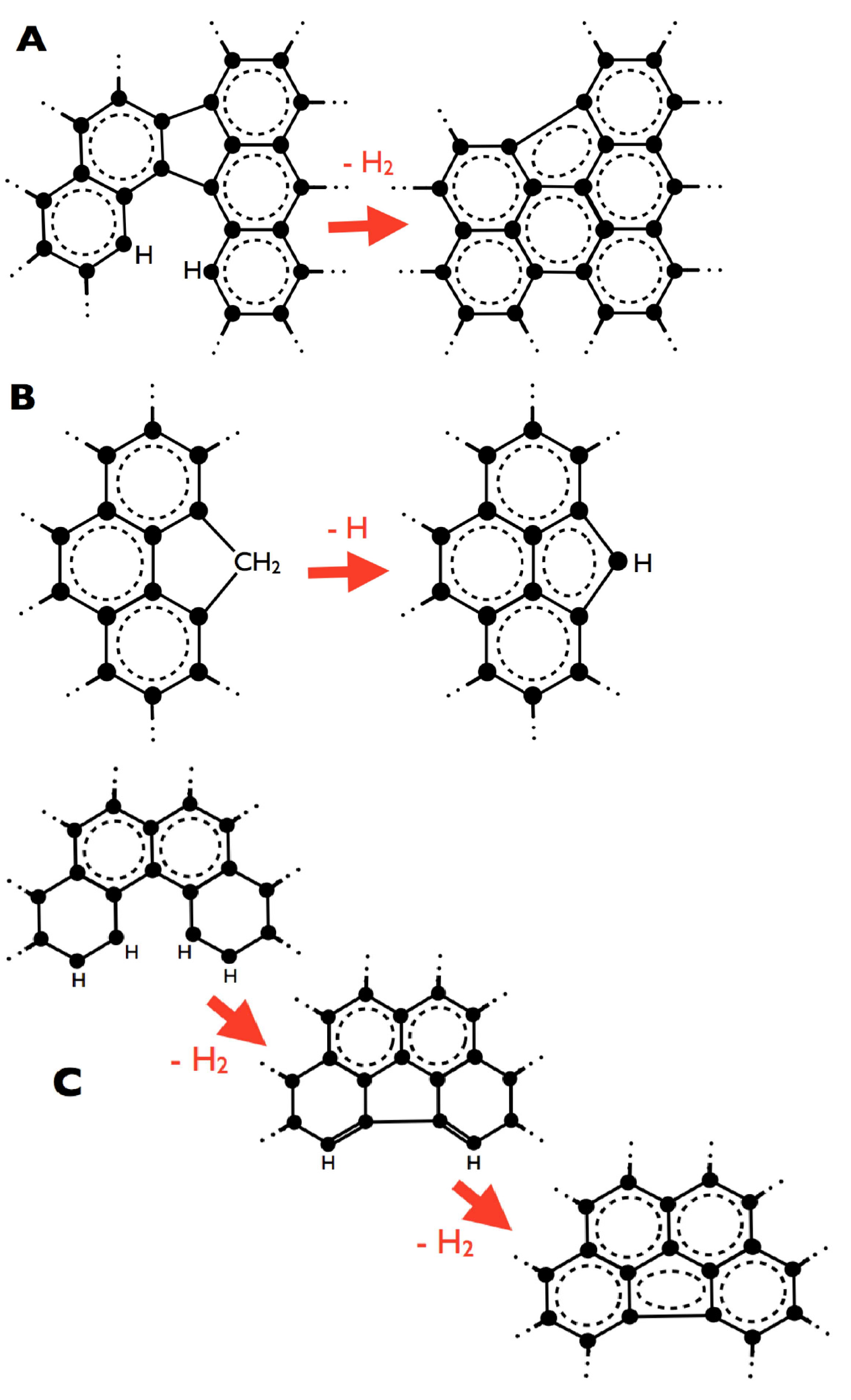}
\caption{A schematic view of three possible routes to aromatic pentagon
   formation (labeled A, B and C), leading to a loss of planarity, in nm-sized a-C:H
   particles via aliphatic and olefinic ring system dehydrogenation. The filled 
   circles represent carbon.}
 \label{fig_schemes}
\end{figure}
%

The aromatisation of the aliphatic bridges to form pentagons can
provide a viable route to fullerene formation in the solid phase via
these ``arophatic'' species.  The key process is the dehydrogenation
of the aliphatic and olefinic ``bridges'' which naturally leads to
aromatic pentagon formation, as shown in Fig.~\ref{fig_schemes} for
small a-C:H particle fragments.
It appears that there is a link between small carbonaceous particle
dehydrogenation in intense UV radiation fields and the formation of
fullerenes, which could explain the observation of fullerenes
in circumstellar regions associated with small hydrocarbon grains
\cite[{\it e.g.},][]{cami10, garcia10, garcia11b, cami11, bsalas12}.

\subsection{Fullerene formation in space}

The mechanism that we propose for fullerene formation in space is 
the following. The starting point is large, nm-sized, H-rich a-C:H
particles containing more than a hundred carbon atoms. UV photolysis
by photons with energy above 10 eV dehydrogenates these grains
\citep{jones12a} and leads to further changes in the molecular
structure -- a process that we call {\bf STIR} ({\bf S}tructural
{\bf T}ransformations in {\bf I}ntense {\bf R}adiation fields). STIR
processing requires constant heating (by photon absorption) to keep
the particle in a vibrationally excited state that promotes structural
transformations leading to the closing of cage-like structures within
the a-C:H nanoparticle.  Amorphous carbon nanoparticles can be
maintained at thermal-equilibrium temperatures of the order of
100-150~K by photon absorption (for H-rich a-C:H particles with band
gaps larger than 1 eV) even at thousands of astronomical units from
the central stars of PNe, due to their low emissivity at the long
wavelengths where they radiate thermally \citep{bsalas12}. However, at
these same distances, almost completely dehydrogenated nanoparticles
have temperatures of only tens of Kelvin, as expected.

\begin{figure}[t]
\centering
\includegraphics*[width=\hsize]{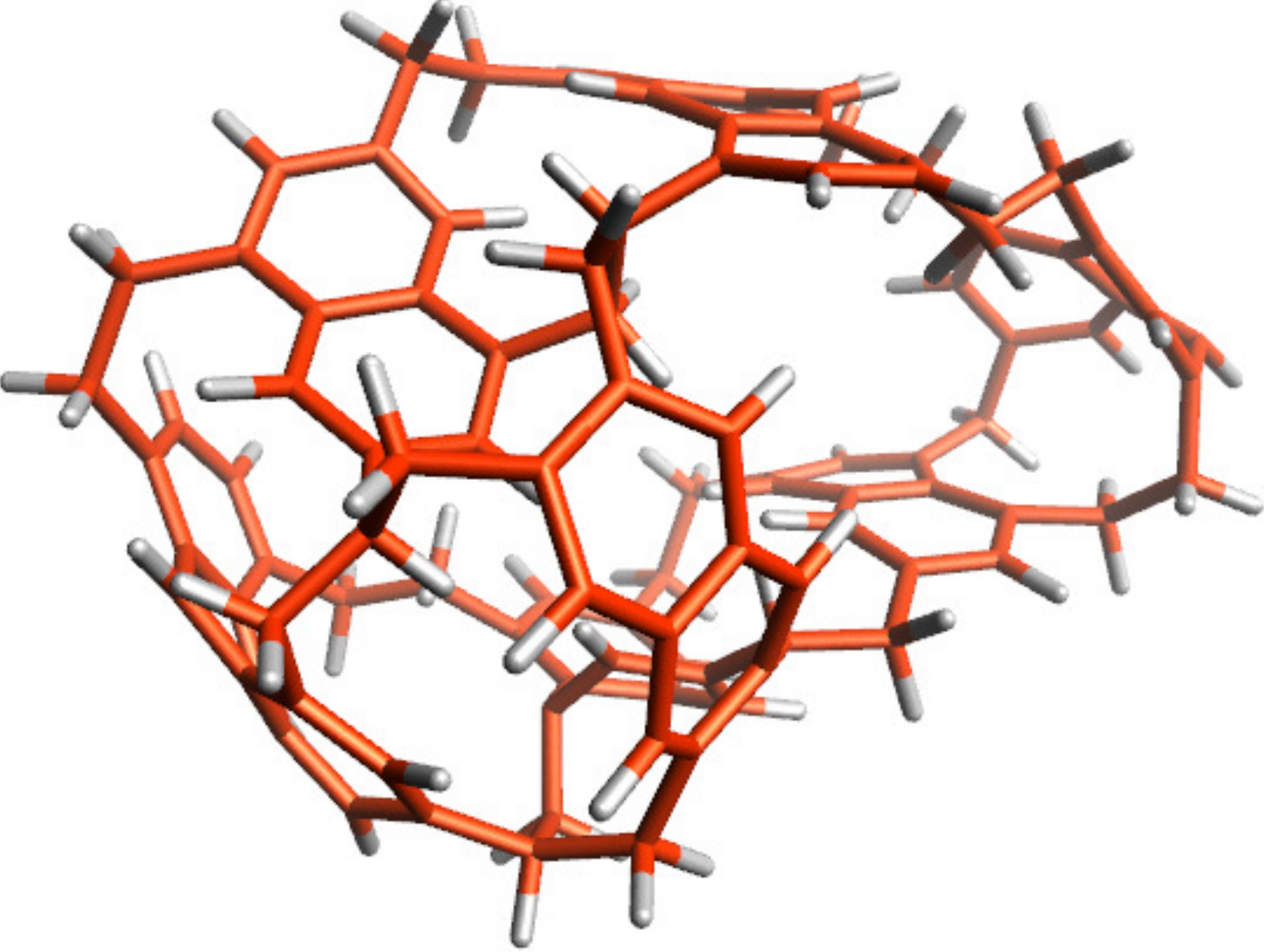}
 \caption{The wrapped version of a large and complex ``arophatic''
   cluster (C$_{104}$H$_{80}$), precursor of cosmic fullerenes. Carbon is
   represented in dark grey, hydrogen in light gray. The color version
   of this figure (carbon in red, hydrogen in gray) is available in
   the electronic edition of the {\it Astrophysical Journal}.}
 \label{fig_arophatics_2}
\end{figure}

\begin{figure}
\centering
\includegraphics[width=\hsize]{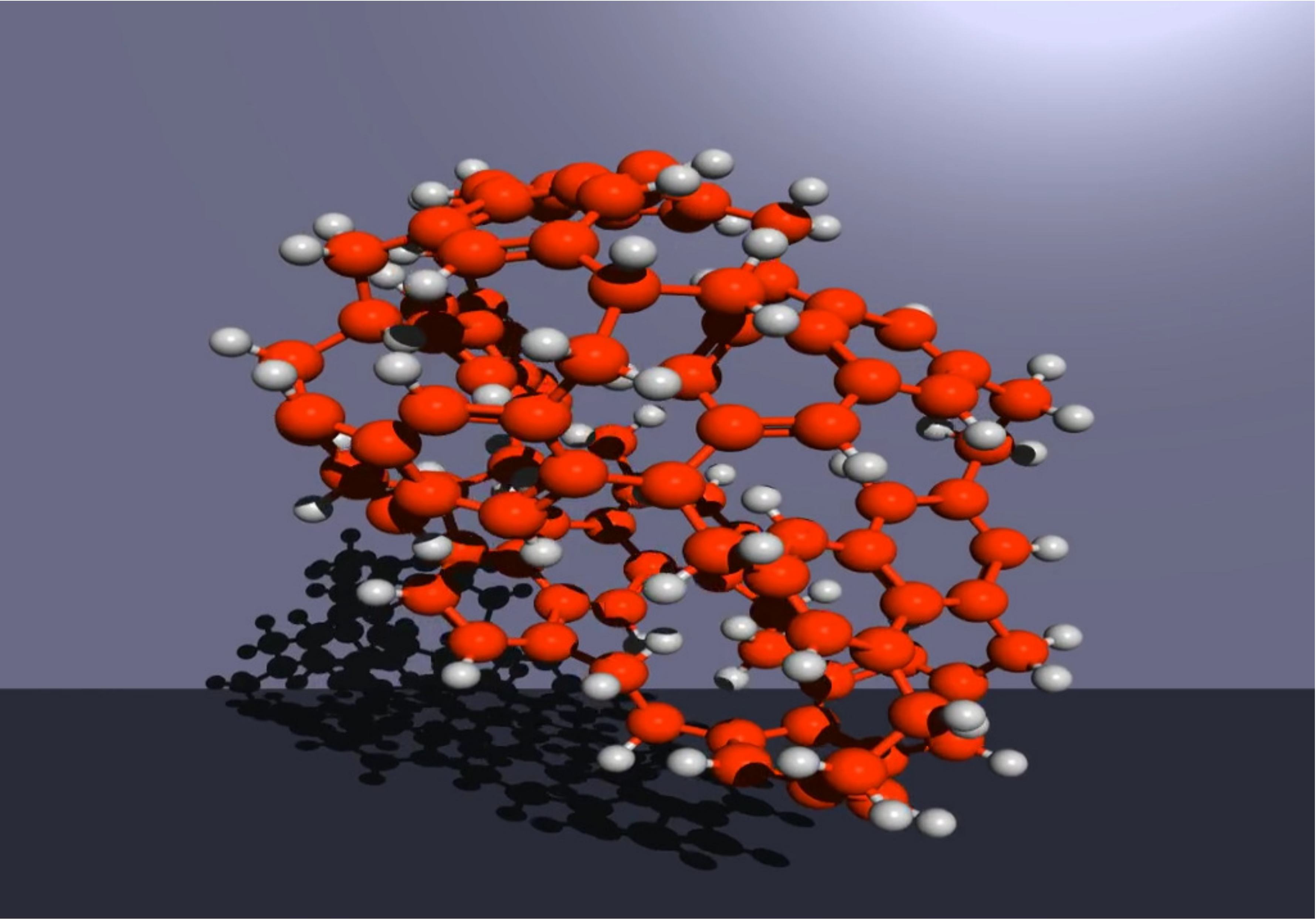}
 \caption{A snap-shot from the movie illustrating the 3-D structure
   of the large and complex ``arophatic'' cluster C$_{104}$H$_{80}$,
   precursor of cosmic fullerenes and schematically represented in
   Fig.~\ref{fig_arophatics_2}. Carbon is in dark grey, hydrogen in
   light gray. The movie, with carbon in red and hydrogen in gray, is
   available in the electronic edition of the {\it Astrophysical Journal}.}
 \label{fig_movie2_shot}
\end{figure}

In order to form fullerenes, the arophatic clusters resulting from
STIR processing must be large,
containing more than a hundred carbon atoms,  
as exemplified in Figs.~\ref{fig_arophatics_2} and
\ref{fig_movie2_shot}.

A key step in the route to fullerenes is the formation of 5-membered
rings. This can occur via dehydrogenation during the
STIR processing of the parent a-C:H or of the emerging arophatic
cluster, in a similar fashion as shown in Fig.~\ref{fig_schemes} for
small a-C:H particles. 
Dehydrogenation is a viable way to introduce
5-membered rings, as indicated by \citet{violi04}. This work shows
that structures with a high degree of curvature result from
dehydrogenation reactions occurring in aromatic hydrocarbon
compounds. These reactions involve the loss of hydrogen with
rearrangement of the structure and ring closure which leads to
five-membered rings, and hence curvature.
%
%
UV irradiation  of  HAC nanoparticles not only causes
dehydrogenation \citep[][and references therein]{jones12a}, 
but also forces the structure to curl up because of
the introduction of pentagonal rings. In fact, the UV-induced
dehydrogenation {\it is} the cause of the formation of 5-membered rings.
Thus, in a single step, dehydrogenation fulfills two essential
conditions for the formation of fullerene molecules: 1) removing
hydrogen atoms and 2) triggering pentagonal ring formation in the
evolving structure.
%
%
%

The fullerene precursor is a big and complex arophatic
cluster, with the aromatic domains being non-co-planar. The 3-D
structure favors the formation of 5-membered aromatic rings, and then
curvature, even in such a big structure, via photo-induced
dehydrogenation. The particle emerging from the dehydrogenation
process is a giant fullerenic cage similar to the ones formed in
simulations \citep{irle06}. The initial, STIR-driven (catastrophic) photolysis and
the following cage closure induced by the formation of pentagonal
rings will result in a highly excited structure.  The cage will
release the excess energy via ejection of C$_{\rm 2}$ molecules,
shrinking down to the islands of stability represented by C$_{\rm 60}$
and C$_{\rm 70}$ \citep{irle06}. The down-sizing occurs through
unimolecular decomposition of the vibrationally excited cage,
resulting in the ejection of carbon molecules. To model the unimolecular
dissociation we adopt an Arrhenius law \citep{tielens05}, which describes the
evolution of the dissociation rate as a function of the temperature of
the dissociating species:
\begin{equation}
\label{Eq:arrhenius_generic}
k_{\rm s}(T) = k_0\,\exp\left(-E_0/kT\right)
\end{equation}
The quantity $k_{\rm s}$ is the rate for carbon ejection measured in
carbon atom lost per second (C/s), $T$ the vibrational temperature of
the fullerenic cage and $k$ the Boltzmann constant. The unimolecular
decomposition is governed by two parameters: the pre-exponential
factor $k_0$ and the dissociation energy $E_0$. To estimate the value
of these two parameters, we use two sets of simulations which follow
the shrinking of giant fullerenic cages at two different temperatures,
2000 and 3000 K, and provide the corresponding dissociation rates.
The average dissociation rate $k_{\rm s}$ is 1.5$\times$10$^{10}$ C/s
and 3$\times$10$^{10}$ C/s for $T$~=~2000 K and $T$~=~3000 K
respectively \citep{zheng07}. We obtain the following values for $k_0$
and $E_0$:
\begin{eqnarray*}
  k_0 & = & 1.2 \times 10^{11}   \,\,\, {\rm C/s} \\
  E_0 & = & 0.36 \,\,\, {\rm eV}
\end{eqnarray*}

The time $t_{\rm s}$ required for a big fullerenic cage to shrink down
to C$_{60}$ is given by:
\begin{equation}
\label{Eq:time_shrink}
t_{\rm s}(T) = N_{\rm ej}/k_{\rm s}(T)
\end{equation}
where $N_{\rm ej}$ is the number of carbons that have to be ejected
from the cage and $k_{\rm s}$ is the ejection rate from
Eq.~\ref{Eq:arrhenius_generic}.  We consider a fullerenic cage
containing 144 carbon atoms, i.e. with size comparable to the size of
the particles considered in simulations \citep{zheng07}, therefore,
$N_{\rm ej}$ = 84. The shrinking timescales for various temperatures
of the initial cage are reported in Table~\ref{Tab:shrink_time}.
%
\begin{table}[t]
\begin{center}
\caption{\label{Tab:shrink_time} Timescale, $t_{\rm s}$, for shrinking of the
  fullerenic cage C$_{144}$ to C$_{60}$.}
\begin{tabular}{c l l}
\hline
\hline
$T$  &  $\phantom{1.8.}k_{\rm s}$  &  $\phantom{1.1}t_{\rm s}$ \\
(K)    & $\phantom{1.}$(C/s)            &    $\phantom{1..}$(s)            \\
\hline
  1000  &  1.8$\times$10$^{9}$  &  4.7$\times$10$^{-8}$  \\
 300         &  1.1$\times$10$^{5}$  &  0.7$\times$10$^{-3}$  \\
 200         &  1.0$\times$10$^{2}$  &  8.4$\times$10$^{-1}$  \\
 100         &  1.0$\times$10$^{-7}$  &  8.4$\times$10$^{8}$    \\
\hline
\end{tabular}
\end{center}
\tablecomments{$T$ is the temperature
 of the cage and $k_{\rm s}$ is the rate for carbon ejection measured in
carbon atom lost per second.}
\end{table}
%
This process requires relatively short time scales: to shrink a large
(144 carbon atoms) fullerene cage to C$_{60}$ requires only about 0.8
s at 200 K, and increases to 27 years at 100 K.

To the best of our knowledge, our extrapolation is the first
attempt to characterize the shrinking of a giant fullerenic cage under
interstellar conditions (e.g., very low temperatures with respect to
the usual laboratory conditions). The value of 0.36 eV that we have
calculated for $E_0$ should be considered as a lower limit, having
been determined under the most favourable conditions for fullerene
formation on Earth, as indicated by simulations (nevertheless the
optimized conditions for industrial production of fullerene are quite
similar). The issue that we are facing here is another example of the
well known problem related to the extrapolation to extraterrestrial
conditions of the results from statistical unimolecular dissociation
theories \citep{tielens05, micelotta10a}. The proper evaluation of
$E_0$ and of the uncertainty associated with it requires the
development of a specific model which will be the subject of a
following paper. Our present scope is to provide a first indication of
the timescales associated with the shrinking process in space. In the
case of PAH dissociation under extraterrestrial conditions, the detailed evaluation of the
dissociation parameter $E_0$ resulted in a factor of 10 uncertainty on the
calculated fragmentation timescales \citep{micelotta10a}. Because of
the similarities between large PAHs and fullerenes
\citep{micelotta10b, micelotta10a}, we can consider a
$10^2$ uncertainty for the fullerene shrinking timescale as a very
conservative assumption. We obtain then a maximum shrinking time of
2670 years at $T$=100 K (the least favourable condition). This is
fully consistent with the age of young PNe where fullerenes have been
observed.

C$_{\rm 60}$ and C$_{\rm 70}$ are the only, and therefore smallest,
C$_{n}$ fullerenes with $n \leq$ 70 satisfying the Isolated Pentagon
Rule (IPR). This implies (almost) spherical, thermodynamically
favored, closed shell structures particularly suited to survive harsh
conditions. The shrinking stops at C$_{60}$ because of the very high
Arrhenius dissociation energy $E_0$ for C$_{2}$ elimination to
C$_{58}$, $E_0 \sim$~10~eV \citep{tomita01}.
The estimated temperatures of a few hundreds of
Kelvin for fullerenes detected in planetary nebulae \citep{cami10,
bsalas12} are far from the approximately 1000 K required to overcome
this decomposition barrier \citep{leifer95}.

Our model proposes the formation of a single C$_{\rm 60}$ (C$_{\rm
70}$) molecule from a single large arophatic cluster. This implies
that the parent a-C:H particles must be nm-sized, and therefore
contain a hundred or more carbon atoms, because of the dissociative
loss of carbon (and hydrogen) atoms that will occur during structural
transformation.  Note that the large arophatic clusters can be further
fragmented by the same radiation field that caused their formation
from the parent a-C:H grain.  Most of the formed arophatics will
fragment into structures other than fullerenes, resulting in a
fullerene formation process that is rather inefficient, consistent
with the low fullerene abundance observed in space so far.

\section{Fullerene observations in planetary nebulae}\label{observations_sec}

The peculiarities of the PNe where fullerenes have been detected are
their intense UV radiation fields, sufficient to sustain
photo-chemical processing of HACs, and the ablation of carbonaceous
material from the dense circumstellar matter formed during their AGB
phase. Thus, in these objects abundant, nm-sized, H-rich, a-C:H dust
is being liberated and undergoing catastrophic UV-photolysis, which
leads to STIR processing.  In Tc 1, C$_{\rm 60}$ and C$_{\rm 70}$ have
been detected far for the central star \citep[8000 AU --
][]{bsalas12}, but even at this distance the UV irradiation is
sufficient to keep the particles in a vibrationally excited state ($T
\sim$ 100-130~K) and to trigger and sustain the STIR processing of the
a-C:H dust and the shrinking of the resulting giant fullerenic cages
down to C$_{\rm 60}$ and C$_{\rm 70}$.

It appears that the young PNe where fullerenes have been observed are
the optimised factories for their formation because their stellar
emission peaks in the required 10 eV photon energy regime. In their
pre-PNe stages these objects do not have sufficient UV photons to
dehydrogenate and heat the dust, and in their later stages the
radiation field is so much harder that dissociative losses will
destroy the H-rich carbonaceous particles before they can re-structure
to form fullerenes.

\section{Conclusions}\label{conclusions_sec}

We studied the formation of fullerenes C$_{\rm 60}$ and C$_{\rm 70}$
in planetary nebulae, for which a satisfactory formation path is still
missing.

In space, the low density of the gas precludes the formation of
fullerene materials following known vaporization or combustion
synthesis routes even on astronomical timescales.

The direct formation of C$_{\rm 60}$ through dissociation-induced
curvature of dehydrogenated PAHs requires a very specific tuning of
the dissociation parameters, and thus appears unlikely.

The scheme for fullerene formation in space that we propose is based
on a top-down approach. The top of the tree is represented by large 
a-C:H/HAC particles, i.e. materials whose presence in space has been
firmly established.

UV photolysis promotes structural transformations in the parent a-C:H
particle with the formation of what we call ``arophatic'' clusters: 3-D,
hollow structures characterized by the presence of aromatic clusters
linked by aliphatic bridging groups. 

UV-induced dehydrogenation of the emerging arophatic cluster (or the
parent a-C:H) introduces pentagonal rings, and hence curvature, in the
evolving structure.

The result is a large, vibrationally-excited fullerenic
cage that shrinks down to the stable molecules C$_{\rm 60}$ and
C$_{\rm 70}$ through emission of C$_2$ groups. The very high energy
barrier for C$_2$ ejection from C$_{\rm 60}$ prevents further
dissociation. 

The young PNe where fullerenes have been observed appear as
the optimised factories for the formation of these molecules. 
In their pre-PNe stages the formation of fullerene-like structure 
is not favoured because these objects lack sufficient UV photons to
dehydrogenate and heat the dust. In their later stages the
radiation field is so much harder that it will destroy H-rich
carbonaceous particles before they can re-structure to form
fullerenes.

\acknowledgments
We would like to thank G. Lavaux for the realization of the movie and
Fig.~\ref{fig_arophatics_2}. E.R.M, J.C. and E.P. acknowledge the
support from the National Science and Engineering Research Council of
Canada (NSERC). J.B-S. wishes to acknowledge the support from a Marie
Curie Intra-European Fellowship within the 7th European Community
Framework Program under project number 272820. This research has made
use of NASA's Astrophysics Data System.



\begin{thebibliography}{52}
\expandafter\ifx\csname natexlab\endcsname\relax\def\natexlab#1{#1}\fi

\bibitem[{{Bernard-Salas} {et~al.}(2012){Bernard-Salas}, {Cami}, {Peeters},
  {Jones}, {Micelotta}, \& {Groenewegen}}]{bsalas12}
{Bernard-Salas}, J., {Cami}, J., {Peeters}, E., {et~al.} 2012, \apj, 757, 41

\bibitem[{{Bern{\'e}} \& {Tielens}(2012)}]{berne12}
{Bern{\'e}}, O., \& {Tielens}, A.~G.~G.~M. 2012, Proceedings of the National
  Academy of Science, 109, 401

\bibitem[{{Cami} {et~al.}(2010){Cami}, {Bernard-Salas}, {Peeters}, \&
  {Malek}}]{cami10}
{Cami}, J., {Bernard-Salas}, J., {Peeters}, E., \& {Malek}, S.~E. 2010,
  Science, 329, 1180

\bibitem[{{Cami} {et~al.}(2011){Cami}, {Bernard-Salas}, {Peeters}, \&
  {Malek}}]{cami11}
{Cami}, J., {Bernard-Salas}, J., {Peeters}, E., \& {Malek}, S.~E. 2011, in IAU
  Symposium, Vol. 280, IAU Symposium, 216--227

\bibitem[{{Cernicharo} {et~al.}(2001){Cernicharo}, {Heras}, {Tielens}, {Pardo},
  {Herpin}, {Gu{\'e}lin}, \& {Waters}}]{cernicharo01}
{Cernicharo}, J., {Heras}, A.~M., {Tielens}, A.~G.~G.~M., {et~al.} 2001, \apjl,
  546, L123

\bibitem[{{Cherchneff} {et~al.}(2000){Cherchneff}, {Le Teuff}, {Williams}, \&
  {Tielens}}]{cherchneff00}
{Cherchneff}, I., {Le Teuff}, Y.~H., {Williams}, P.~M., \& {Tielens},
  A.~G.~G.~M. 2000, \aap, 357, 572

\bibitem[{{Chung} \& {Violi}(2011)}]{chung11}
{Chung}, S.-H., \& {Violi}, A. 2011, Proc. Combust. Inst., 33, 693

\bibitem[{Chuvilin {et~al.}(2010)Chuvilin, Kaiser, Bichoutskaia, Besley, \&
  Khlobystov}]{chuvilin10}
Chuvilin, A., Kaiser, U., Bichoutskaia, E., Besley, N.~A., \& Khlobystov, A.~N.
  2010, Nature Chemistry, 2, 450

\bibitem[{{Clayton} {et~al.}(2011){Clayton}, {De Marco}, {Whitney}, {Babler},
  {Gallagher}, {Nordhaus}, {Speck}, {Wolff}, {Freeman}, {Camp}, {Lawson},
  {Roman-Duval}, {Misselt}, {Meade}, {Sonneborn}, {Matsuura}, \&
  {Meixner}}]{clayton11}
{Clayton}, G.~C., {De Marco}, O., {Whitney}, B.~A., {et~al.} 2011, \aj, 142, 54

\bibitem[{Dunk {et~al.}(2012)Dunk, Kaiser, Hendrickson, Quinn, Ewels,
  Nakanishi, Sasaki, Shinohara, Marshall, \& Kroto}]{dunk12}
Dunk, P.~W., Kaiser, N.~K., Hendrickson, C.~L., {et~al.} 2012, Nat. Commun., 3,
  855

\bibitem[{{Evans} {et~al.}(2012){Evans}, {van Loon}, {Woodward}, {Gehrz},
  {Clayton}, {Helton}, {Rushton}, {Eyres}, {Krautter}, {Starrfield}, \&
  {Wagner}}]{evans12}
{Evans}, A., {van Loon}, J.~T., {Woodward}, C.~E., {et~al.} 2012, \mnras, 421,
  L92

\bibitem[{{Ferrari} \& {Robertson}(2004)}]{2004PhilTransRSocLondA..362.2477F}
{Ferrari}, A.~C., \& {Robertson}, J. 2004, Phil. Trans. R. Soc. Lond. A, 362,
  2477

\bibitem[{{Gadallah} {et~al.}(2011){Gadallah}, {Mutschke}, \&
  {J{\"a}ger}}]{gadallah11}
{Gadallah}, K.~A.~K., {Mutschke}, H., \& {J{\"a}ger}, C. 2011, \aap, 528, A56

\bibitem[{{Garc{\'{\i}}a-Hern{\'a}ndez}
  {et~al.}(2011{\natexlab{a}}){Garc{\'{\i}}a-Hern{\'a}ndez}, {Kameswara Rao},
  \& {Lambert}}]{garcia11a}
{Garc{\'{\i}}a-Hern{\'a}ndez}, D.~A., {Kameswara Rao}, N., \& {Lambert}, D.~L.
  2011{\natexlab{a}}, \apj, 729, 126

\bibitem[{{Garc{\'{\i}}a-Hern{\'a}ndez}
  {et~al.}(2010){Garc{\'{\i}}a-Hern{\'a}ndez}, {Manchado},
  {Garc{\'{\i}}a-Lario}, {Stanghellini}, {Villaver}, {Shaw}, {Szczerba}, \&
  {Perea-Calder{\'o}n}}]{garcia10}
{Garc{\'{\i}}a-Hern{\'a}ndez}, D.~A., {Manchado}, A., {Garc{\'{\i}}a-Lario},
  P., {et~al.} 2010, \apjl, 724, L39

\bibitem[{{Garc{\'{\i}}a-Hern{\'a}ndez}
  {et~al.}(2011{\natexlab{b}}){Garc{\'{\i}}a-Hern{\'a}ndez}, {Iglesias-Groth},
  {Acosta-Pulido}, {Manchado}, {Garc{\'{\i}}a-Lario}, {Stanghellini},
  {Villaver}, {Shaw}, \& {Cataldo}}]{garcia11b}
{Garc{\'{\i}}a-Hern{\'a}ndez}, D.~A., {Iglesias-Groth}, S., {Acosta-Pulido},
  J.~A., {et~al.} 2011{\natexlab{b}}, \apjl, 737, L30

\bibitem[{{Gielen} {et~al.}(2011){Gielen}, {Cami}, {Bouwman}, {Peeters}, \&
  {Min}}]{gielen11}
{Gielen}, C., {Cami}, J., {Bouwman}, J., {Peeters}, E., \& {Min}, M. 2011,
  \aap, 536, A54

\bibitem[{{Goto} {et~al.}(2003){Goto}, {Gaessler}, {Hayano}, {Iye}, {Kamata},
  {Kanzawa}, {Kobayashi}, {Minowa}, {Saint-Jacques}, {Takami}, {Takato}, \&
  {Terada}}]{goto03}
{Goto}, M., {Gaessler}, W., {Hayano}, Y., {et~al.} 2003, \apj, 589, 419

\bibitem[{{Grishko} \& {Duley}(2000)}]{2000ApJ...543L..85G}
{Grishko}, V.~I., \& {Duley}, W.~W. 2000, \apjl, 543, L85

\bibitem[{{Grishko} {et~al.}(2001){Grishko}, {Tereszchuk}, {Duley}, \&
  {Bernath}}]{grishko01}
{Grishko}, V.~I., {Tereszchuk}, K., {Duley}, W.~W., \& {Bernath}, P. 2001,
  \apjl, 558, L129

\bibitem[{{Howard} {et~al.}(1991){Howard}, {McKinnon}, {Makarovsky}, {Lafleur},
  \& {Johnson}}]{howard91}
{Howard}, J.~B., {McKinnon}, J.~T., {Makarovsky}, Y., {Lafleur}, A.~L., \&
  {Johnson}, M.~E. 1991, \nat, 352, 139

\bibitem[{Irle {et~al.}(2006)Irle, Zheng, Wang, \& Morokuma}]{irle06}
Irle, S., Zheng, G., Wang, Z., \& Morokuma, K. 2006, The Journal of Physical
  Chemistry B, 110, 14531, pMID: 16869552

\bibitem[{{J{\"a}ger} {et~al.}(2009){J{\"a}ger}, {Huisken}, {Mutschke},
  {Jansa}, \& {Henning}}]{jager09}
{J{\"a}ger}, C., {Huisken}, F., {Mutschke}, H., {Jansa}, I.~L., \& {Henning},
  T. 2009, \apj, 696, 706

\bibitem[{{Jones}(2012{\natexlab{a}})}]{jones12c}
{Jones}, A.~P. 2012{\natexlab{a}}, \aap, 540, A1+

\bibitem[{{Jones}(2012{\natexlab{b}})}]{jones12a}
---. 2012{\natexlab{b}}, \aap, 540, A2+

\bibitem[{{Jones}(2012{\natexlab{c}})}]{jones12b}
---. 2012{\natexlab{c}}, \aap, 542, A98+

\bibitem[{{Kr{\"a}tschmer} {et~al.}(1990{\natexlab{a}}){Kr{\"a}tschmer},
  {Fostiropoulos}, \& {Huffman}}]{kratschmer90a}
{Kr{\"a}tschmer}, W., {Fostiropoulos}, K., \& {Huffman}, D.~R.
  1990{\natexlab{a}}, Chem. Phys. Lett., 170, 167

\bibitem[{{Kr{\"a}tschmer} {et~al.}(1990{\natexlab{b}}){Kr{\"a}tschmer},
  {Lamb}, {Fostiropoulos}, \& {Huffman}}]{kratschmer90b}
{Kr{\"a}tschmer}, W., {Lamb}, L.~D., {Fostiropoulos}, K., \& {Huffman}, D.~R.
  1990{\natexlab{b}}, \nat, 347, 354

\bibitem[{{Kroto} {et~al.}(1985){Kroto}, {Heath}, {Obrien}, {Curl}, \&
  {Smalley}}]{kroto85}
{Kroto}, H.~W., {Heath}, J.~R., {Obrien}, S.~C., {Curl}, R.~F., \& {Smalley},
  R.~E. 1985, \nat, 318, 162

\bibitem[{{Kwok}(2009)}]{kwok09}
{Kwok}, S. 2009, International Journal of Astrobiology, 8, 161

\bibitem[{{Leifer} {et~al.}(1995){Leifer}, {Goodwin}, {Anderson}, \&
  {Anderson}}]{leifer95}
{Leifer}, S.~D., {Goodwin}, D.~G., {Anderson}, M.~S., \& {Anderson}, J.~R.
  1995, \prb, 51, 9973

\bibitem[{{Micelotta} {et~al.}(2010{\natexlab{a}}){Micelotta}, {Jones}, \&
  {Tielens}}]{micelotta10b}
{Micelotta}, E.~R., {Jones}, A.~P., \& {Tielens}, A.~G.~G.~M.
  2010{\natexlab{a}}, \aap, 510, A36+

\bibitem[{{Micelotta} {et~al.}(2010{\natexlab{b}}){Micelotta}, {Jones}, \&
  {Tielens}}]{micelotta10a}
---. 2010{\natexlab{b}}, \aap, 510, A37+

\bibitem[{{Nano-C}(2004)}]{nanoC04}
{Nano-C}. 2004, {Masters of the Flame: Industrial Production of Fullerenes
  Becomes a Reality} (http://www.nano-c.com/technologies.as)

\bibitem[{{Pardo} {et~al.}(2007){Pardo}, {Cernicharo}, {Goicoechea},
  {Gu{\'e}lin}, \& {Asensio Ramos}}]{pardo07}
{Pardo}, J.~R., {Cernicharo}, J., {Goicoechea}, J.~R., {Gu{\'e}lin}, M., \&
  {Asensio Ramos}, A. 2007, \apj, 661, 250

\bibitem[{{Peeters} {et~al.}(2012){Peeters}, {Tielens}, {Allamandola}, \&
  {Wolfire}}]{peeters12}
{Peeters}, E., {Tielens}, A.~G.~G.~M., {Allamandola}, L.~J., \& {Wolfire},
  M.~G. 2012, \apj, 747, 44

\bibitem[{{Petrie} {et~al.}(2003){Petrie}, {Stranger}, \&
  {Duley}}]{2003ApJ...594..869P}
{Petrie}, S., {Stranger}, R., \& {Duley}, W.~W. 2003, \apj, 594, 869

\bibitem[{{Rauls} \& {Hornek{\ae}r}(2008)}]{2008ApJ...679..531R}
{Rauls}, E., \& {Hornek{\ae}r}, L. 2008, \apj, 679, 531

\bibitem[{{Roberts} {et~al.}(2012){Roberts}, {Smith}, \& {Sarre}}]{roberts12}
{Roberts}, K.~R.~G., {Smith}, K.~T., \& {Sarre}, P.~J. 2012, \mnras, 421, 3277

\bibitem[{{Rubin} {et~al.}(2011){Rubin}, {Simpson}, {O'Dell}, {McNabb},
  {Colgan}, {Zhuge}, {Ferland}, \& {Hidalgo}}]{rubin11}
{Rubin}, R.~H., {Simpson}, J.~P., {O'Dell}, C.~R., {et~al.} 2011, \mnras, 410,
  1320

\bibitem[{{Scott} \& {Duley}(1996)}]{1996ApJ...472L.123S}
{Scott}, A., \& {Duley}, W.~W. 1996, \apjl, 472, L123+

\bibitem[{{Scott} {et~al.}(1997){Scott}, {Duley}, \&
  {Pinho}}]{1997ApJ...489L.193S}
{Scott}, A., {Duley}, W.~W., \& {Pinho}, G.~P. 1997, \apjl, 489, L193+

\bibitem[{{Sellgren} {et~al.}(2010){Sellgren}, {Werner}, {Ingalls}, {Smith},
  {Carleton}, \& {Joblin}}]{sellgren10}
{Sellgren}, K., {Werner}, M.~W., {Ingalls}, J.~G., {et~al.} 2010, \apjl, 722,
  L54

\bibitem[{{Smith}(1984)}]{1984JAP....55..764S}
{Smith}, F.~W. 1984, Journal of Applied Physics, 55, 764

\bibitem[{Taylor {et~al.}(1990)Taylor, Hare, Abdul-Sada, \& Kroto}]{taylor90}
Taylor, R., Hare, J.~P., Abdul-Sada, A.~K., \& Kroto, H.~W. 1990, J. Chem.
  Soc.{,} Chem. Commun., 1423

\bibitem[{{Tielens}(2005)}]{tielens05}
{Tielens}, A.~G.~G.~M. 2005, {The Physics and Chemistry of the Interstellar
  Medium} (Cambridge Univ. Press, Cambridge)

\bibitem[{{Tomita} {et~al.}(2001){Tomita}, {Andersen}, {Gottrup}, {Hvelplund},
  \& {Pedersen}}]{tomita01}
{Tomita}, S., {Andersen}, J.~U., {Gottrup}, C., {Hvelplund}, P., \& {Pedersen},
  U.~V. 2001, Phys. Rev. Lett., 87, 073401

\bibitem[{{Ugarte}(1992)}]{ugarte92}
{Ugarte}, D. 1992, \nat, 359, 707

\bibitem[{{Violi}(2004)}]{violi04}
{Violi}, A. 2004, Combustion and Flame, 139, 279

\bibitem[{{Zhang} \& {Kwok}(2011)}]{zhangkwok11}
{Zhang}, Y., \& {Kwok}, S. 2011, \apj, 730, 126

\bibitem[{Zheng {et~al.}(2005)Zheng, Irle, \& Morokuma}]{zheng05}
Zheng, G., Irle, S., \& Morokuma, K. 2005, The Journal of Chemical Physics,
  122, 014708

\bibitem[{{Zheng} {et~al.}(2007){Zheng}, {Wang}, {Irle}, \&
  {Morokuma}}]{zheng07}
{Zheng}, G., {Wang}, Z., {Irle}, S., \& {Morokuma}, K. 2007, J. Nanosci.
  Nanotechnol., 7, 1662
\end{thebibliography}

\appendix

\section{Timescale for fullerene condensation in space}

We report here the details of the calculation of the scaling rule
(Eq.~\ref{scaling_eq}), which relates the time $\tilde{\tau}$ required
for fullerene condensation with the initial density $n$ of the carbon
gas. Consider an ensemble of $N_{\rm p}$ particles that can occupy
$N_{\rm v}$ cells. The probability of having one
particle in one cell is $p = N_{\rm p}/N_{\rm v} \ll 1$. The
probability $P$ of having two particles in the same cell at the same
time is described by the Poisson statistics:
\begin{equation}
\label{poisson_eq}
  P(2; p) = \frac{p^2}{2}\,e^{-p}
\end{equation}
We have that $N_{\rm v} \equiv V/\delta v$, where $V$ is the total
volume and $\delta v$ is the volume of a single cell. We can then
rewrite the single occupancy probability as $p = (N_{\rm p}/V)\,\delta
v = n\,\delta v$.

The probability $\tilde{P}$ of having interaction between two
particles at time $N_{\rm t}$ can be written as:
\begin{equation}
\label{tilde1_eq}
  \tilde{P}({\rm int}; N_{\rm t}) = (1 - \lambda)^{N_{\rm t}-1}\,\lambda
\end{equation}
where $N_{\rm t} \simeq t/\tau$, $t$ being the time and $\tau$ the
characteristic time of a single interaction (depending on the nature
of the interaction itself). The term $\lambda$ represents
the probability of interaction between two particles at a generic time
$t$. The probability of having interaction between two
particles up to the time $N_{\rm t}$ is then given by the following
equation:
\begin{equation}
\label{tilde2_eq}
  \tilde{P}({\rm int}; \leq N_{\rm t}) = \sum_{k=0}^{N_{\rm
        t}-1} (1 - \lambda)^{k}\,\lambda = 1 - (1 - \lambda)^{N_{\rm t}}
\end{equation}
If $\lambda$ is small, we can rewrite Eq.~\ref{tilde2_eq} as:
\begin{equation}
\label{tilde3_eq}
  \tilde{P}({\rm int}; \leq N_{\rm t}) \simeq 1 - e^{-\lambda\,N_{\rm
      t}} \simeq 1 -e^{-\lambda\,(t/\tau)}
\end{equation}
The probability $\lambda$ can be written in the following way:
\begin{equation}
\label{lambda_eq}
  \lambda = A\,P(2; p) \simeq A(n\,\delta v)^2
\end{equation}
The term $A$ includes the characteristics intrinsic to the interaction
(which we do not need to specify) while $P(2; p)$ is the probability
from Eq.~\ref{poisson_eq}. The approximation resulting in the right-hand term of the
equation is valid for $p \rightarrow 0$. Combining Eqs.~\ref{tilde3_eq} and \ref{lambda_eq},
with $A\,\delta v^2 = B$, we obtain
\begin{equation}
\label{tilde4_eq}
  \tilde{P}({\rm int}; \leq N_{\rm t}) \simeq 1 - e^{A(n\,\delta
    v)^2\,(t/\tau)} \simeq 1 - e^{B\,n^2 t/\tau}
\end{equation}

From Eq.~\ref{tilde4_eq} we can derive the timescale for condensation:
\begin{equation}
\label{timescale_eq}
  \tilde{\tau} = \frac{\tau}{n^2 B}
\end{equation}
Because both $\tau$ and $B$ are independent from the initial particle
density $n$, we can write the following scaling rule
\begin{equation}
\label{scalingApp_eq}
  \frac{\tilde{\tau_2}}{\tilde{\tau_1}} = \frac{n_1^2}{n_2^2}\,\,
  \Rightarrow \,\, \tilde{\tau_2} = \frac{\tilde{\tau_1} n_1^2}{n_2^2}
\end{equation}
where $\tilde{\tau_1}$ and $\tilde{\tau_2}$ are the condensation
timescales corresponding to densities $n_1$ and $n_2$ respectively.


\end{document}